\documentclass[10pt,twocolumn]{article}
\usepackage{epsf}
\def\1ad{\mbox{\normalsize $^1$}}
\def\2ad{\mbox{\normalsize $^2$}}
\def\3ad{\mbox{\normalsize $^3$}}
\def\4ad{\mbox{\normalsize $^4$}}
\def\5ad{\mbox{\normalsize $^5$}}
\def\6ad{\mbox{\normalsize $^6$}}
\def\7ad{\mbox{\normalsize $^7$}}
\def\8ad{\mbox{\normalsize $^8$}}

\def\dj{\hbox{d\kern-0.347em \vrule width 0.3em height 1.252ex depth
-1.21ex \kern 0.051em}}

\def\lim{\mbox{{\bf L}} }

\newcommand{\be}{\begin{equation}}
\newcommand{\ee}{\end{equation}}
\newcommand{\ben}{\begin{equation*}}
\newcommand{\een}{\end{equation*}}
\newcommand{\ba}{\begin{eqnarray}}
\newcommand{\ea}{\end{eqnarray}}
\newcommand{\ban}{\begin{eqnarray*}}
\newcommand{\ean}{\end{eqnarray*}}
\newcommand{\brr}{\begin{array}}
\newcommand{\err}{\end{array}}
\newcommand{\bc}{\begin{center}}
\newcommand{\ec}{\end{center}}

\begin{document}

{\large
\newcommand{\sheptitle}
{STRING THEORY}
\newcommand{\shepauthora}
{{\sc
 J.L.F.~Barb\'on}
\footnote[1]{On leave from
Departamento de F\'{\i}sica
de Part\'{\i}culas da Universidade de Santiago de Compostela, Spain.}
}

\newcommand{\shepaddressa}
{\sl
Department of Physics, Theory Division, CERN \\
 CH-1211 Geneva 23, Switzerland \\
{\tt barbon@cern.ch}}



\newcommand{\shepabstract}
{
This is a rendering of a review talk on the state of String Theory, given at
the EPS-2003  Conference, intended for a wide audience of experimental and theoretical
physicists. It emphasizes general ideas rather than technical aspects.
}

\begin{titlepage}
\begin{flushright}
{CERN-PH-TH/2004-048 \\
{\tt hep-th/0404188}}

\end{flushright}
\vspace{1in}
\begin{center}
{\large{\bf \sheptitle}}
\bigskip\bigskip \\ \shepauthora \\ \mbox{} \\ {\it \shepaddressa} \\
\vspace{1in}

{\bf Abstract} \bigskip \end{center} \setcounter{page}{0}
 \shepabstract
\vspace{2in}
\begin{flushleft}
CERN-PH-TH/2004-048\\
\today
\end{flushleft}


\end{titlepage}
}

\newpage


\setcounter{equation}{0}

\noindent

String theory is the dominant framework for the construction of 
a unified quantum theory including gravity. In the last decade, the
theory underwent a major conceptual revolution whose consequences
are still unfolding. In this article I  review a selected number of
recent research directions, presented against the background of
various well established theoretical results.   

The choice of topics is fairly subjective, and no attempt was made at collecting  
a complete reference list. General reviews of the subject with extensive
lists of references can be found in
\cite{gsw, pol, magoo}.

\section{What do we know?}

\noindent

Perhaps the most remarkable  property of string theory   is the emergence of gravity
from a purely mechanical model. Starting from  flat space-time, quantizing  
a  relativistic closed string   
without internal structure (i.e. a mathematically {\it thin} string)
 yields a quantum
model of the graviton.  
This means that one finds universally a massless mode of spin 2, which couples
at low energies according to General Relativity (GR). At the same time, the interaction
is soft at high energies, simply because highly energetic strings tend to grow in size, 
which smears the interactions. 

String theory provides the only known 
quantum model of the graviton that is consistent at all energies, and hence it is
the starting point of a theory of quantum gravity. In this sense,  we may regard
closed strings
as the
 natural microscopic ``excitations" of quantum space-time. In perturbation theory, any
string theory is characterized by a fundamental mass scale, $M_s$, that sets
the intrinsic tension of the string. It is also characterized by a dimensionless
coupling constant, $g_s <1$, and together they determine the Planck length (or
Newton's constant) according
to a  relation of the form:  
\begin{equation}
G_{\rm N}  \sim \ell_{\rm P}^{\;2} \sim g_s^{\;2} \;\ell_s^{\;2}\;,
\end{equation}
where $\ell_s = M_s^{-1}$ is the fundamental string length scale and
$\ell_{\rm P}$ is the Planck length. This formula
is  analogous to the expression for the Fermi constant of weak interactions
in terms of the weak Yang--Mills coupling and the mass of the vector bosons:
\begin{equation}
G_{\rm F} \sim g_W^{\;2} \;M_W^{-2}\;.
\end{equation}
In fact, the logical status of both formulas is very similar, i.e. string
theory is a {\it perturbative} smearing of the gravitational interaction.

String perturbation theory is a very tight formalism with a high
degree of self-consistency. Heuristically, the peculiar properties
of perturbative strings are related to the topological character of
the elementary string interactions; once the free propagation of
a string is known (the free spectrum), the interactions are determined
to all orders in the string coupling $g_s$.

Along with the graviton at the massless level,  string models feature  
scalar, fermion, and gauge field excitations, the building blocks of
the Standard Model (SM). Quasi-realistic models can be constructed at the 
qualitative level. This property has led to an interpretation of string
theories as unification models including gravity. In fact, they incorporate
many theoretical schemes  that predate them, such as
supersymmetry and extra dimensions of space-time, and came to dominate the
art of speculative model building for the last two decades.  

\subsection{Supersymmetry  and stability} 

\noindent
  
In many models, such as the simplest bosonic string, there are modes
with $M^2 <0$, i.e. tachyons that signify a dynamical instability of 
the space-time. The only {\it generic} cure known for such instabilities is
the assumption of space-time supersymmetry. In supersymmetric models,
the spectrum in Minkowski space  admits the action of a superalgebra
\begin{equation}
\{\,Q, Q\,\} \sim H + \dots, 
\end{equation} 
where $H$ denotes the Hamiltonian, and the vacuum is left 
 invariant $Q\,|{\rm vac}\,\rangle=0$. Under such conditions, the
operator $M^2$ is non-negative. This stability property
is robust, in the sense  
that particular examples with approximate supersymmetry are free from tachyons,
although non-zero  tadpoles may induce mild instabilities.
Approximate supersymmetry means that the scale of its breaking
$M_{b}$ is small with respect to the string scale $M_{b} \ll M_s$. In other words,
low-energy supersymmetry appears as a ``technical" requirement to
ensure  space-time's local stability.

These considerations about the role of supersymmetry in the local stability of
space-time are based on experience with concrete models. 
Unfortunately, it has not been possible to  
prove a general theorem, much less   predict the numerical
value of the ratio $M_{b} /M_s$. Hence, while the current 
understanding of string theory relies heavily on supersymmetry, we cannot
quite say that $M_b$ is as low as  the TeV scale. 
In any case, the  fact that $M_b \ll M_s$ is preferred from general arguments is
 certainly suggestive of this possibility.

\subsection{Vacuum structure and duality}

\noindent

A high degree of supersymmetry does play a pivotal role in the global structure
of known string models. If the number of independent supercharges ranges between
8 and 32 the string models come in continuous families. The parameters of these
vacua are called {\it moduli} and they are interpreted in space-time as massless
scalar fields. In extreme regions of these {\it moduli spaces} of vacua,  
space-time can be geometrically interpreted as the product of $d$-dimensional
 Minkowski space-time times
a compact manifold: ${\bf R}^d \times K$.
 Typically the dimension  $d_K$ of 
the compact 
 smooth manifold is such that $d_K + d$ is either 10 or 11.  

The low-energy effective theories on
 these manifolds of vacua are supergravity theories
on ${\bf R}^d$ with scalar fields (the moduli) that are interpreted as 
the geometrical parameters of $K$ (size and shape). The strength of the string
coupling can also be understood as one such modulus, the dilaton $\phi$, so that
we can write an equation in all weakly-coupled string models that relates 
the expectation value of the dilaton with the coupling: 
$g_s = \exp \,\langle \phi \rangle$.

Therefore, we find continuous moduli spaces of vacua with extreme regions
 featuring a weakly-coupled description in terms of
string perturbation theory (for $g_s \ll 1$,) and/or low-energy supergravity
for $\ell_K \gg \ell_s,\;\ell_{\rm P}$, in terms of a characteristic
length scale $\ell_K$ of the compact manifold $K$.    

One of the great leaps forward in the 90's was the recognition that different
string theories with {\it a priori} unrelated perturbative expansions could actually
be mapped into one another by a discrete 
non-perturbative symmetry called {\it duality}, a generalization of known
duality symmetries in electrodynamics and statistical mechanics. For
example, the strong coupling limit of ten-dimensional heterotic strings with
gauge group $SO(32)$ turns out to be the weakly-coupled ten-dimensional type-I
superstring theory. Since the type-I theory is a model of {\it open} unoriented
strings and the heterotic model contains closed oriented strings, the two
theories could not be more different at a perturbative level. Yet, they are
related by a mapping of the form 
$$
g_s^{(H)} = 1/g_s^{(I)}\;.
$$ 
Another famous example is the emergence of an eleven-dimensional vacuum
with Poincar\'e symmetry as the strong coupling limit of ten-dimensional
type-IIA strings. 

Under these dualities, perturbative modes of one theory are transformed into
non-perturbative states of another, such as solitons with mass proportional
to $1/g_s$ or $1/g_s^2$. These solitons  are visible in the supergravity
Lagrangian as multidimensional generalizations of extremal black holes called
$p$-branes.    Particular states can be followed from weak to strong coupling, 
when they are protected by supersymmetry, according to
 the so-called BPS phenomenon \cite{bps}.

With many supercharges, it is possible that some finite energy states 
$|\psi_{\rm BPS}
\rangle$ are
annihilated by a subset of the supercharges that leave the vacuum invariant.
We refer to such charges as {\it unbroken}, $Q_u$, and they satisfy 
$$
Q_u \,|\psi_{\rm BPS}  
\rangle =0\;.
$$  
When these states admit a semiclassical description as solitons, the broken
charges (the rest of them) generate Goldstone 
fermions that serve as fermionic collective
coordinates for the low-energy dynamics of the soliton. At any rate, the 
dimensionality of the corresponding unitary representation is smaller than that
of generic states, owing to the vanishing of the $Q_u$ on these states. This
dimensionality being a discrete parameter,
 it cannot change by continuous deformations
such as the variation of $g_s$ or any other modulus.
 The result is that the
BPS states can be
 followed around the supermoduli space and give information about  
its global structure, i.e. we can literally build a discrete ``skeleton" of
this moduli space.
 In this way very complicated moduli spaces with an action
of large duality groups can be unravelled.  

With 4 supercharges on the vacuum (the equivalent of ${\cal N} =1$ in
four dimensions) we can incorporate interesting features such as
chiral fermion spectra at low energies. Exact ${\cal N}=1$ vacua
are expected to be generically 
{\it isolated}, with no moduli, a most interesting fact,  since moduli fields are
always problematic for phenomenological models. 
 Unfortunately, at the level of
 perturbation theory in $g_s$ or $\ell_s /\ell_K$ the ${\cal N}=1$ vacua
still come in moduli spaces, to be lifted only at the {\it non-perturbative
} level. This task has been historically difficult for technical reasons
and, to  date,
  no exact, isolated ${\cal N}=1$ vacuum could be studied in any detail.

All the supersymmetric vacua mentioned so far contain a factor of Minkowski
space. If we consider vacua with asymptocally {\it negative} curvature in 
Lorentzian-signature factors,
we find geometries of the form ${\rm AdS} \times K$ with
AdS denoting Anti-de Sitter space-time and $K$ a suitable compact 
manifold. These vacua are also isolated and admit a non-perturbative
description in terms of the so-called AdS/CFT duality. 

Finally, vacua with no supersymmetry remain largely 
inaccessible to precise theoretical analysis. 

\subsection{Impurities in space-time: singularities and branes}

\noindent

Each point in a supermoduli space of vacua represents a  space-time.  
In  regions where supergravity is a good approximation, there is a
geometrical description of the form ${\bf R}^d \times K$, with $K$ a 
smooth complex manifold with appropriate holonomy group and metric properties  
(Calabi--Yau, K3, G2, etc). However, as the moduli are varied, $K$ may develop
a variety of geometrical  singularities whose physics may 
escape the effective supergravity description. Thus, the study of the structure of
the supermoduli space is largely a question of singularity resolution. Since
string theory is primarily a theory of gravity, the resolution of singularities
is one of the basic problems that must solve.

It turns out that singularities on the appropriate (supersymmetric) manifolds $K$
can be classified to a large extent and analyzed in quite physical terms. The
various known mechanisms of singularity resolution in string theory are always associated
to the emergence of extra light degrees of freedom localized at the singular locus.
These extra light modes are often of topological nature. They might arise at a
purely perturbative level, such as light winding modes at small circles or conical
singularities in orbifolds, or they might be of non-perturbative origin, such as
various wrapped solitonic branes. In many cases, the localized light modes are BPS-protected
and we can write an exact low-energy effective theory that governs their dynamics.

The result is a physical resolution of the singularity when taking into account the
dynamics of the light modes. Sometimes the singularity is just smoothed out by
stringy fuzziness, such as the conical singularities of orbifolds. In other situations,
a similar looking conical singularity (the conifold) develops a new branch of
space-time with nontrivial topological transitions between different manifolds. The
large variety and richness of dynamical resolution of singularities has turned this
problem into more of an art than a craft, the main limitation being the restriction
to supersymmetric types of singularities.

The most interesting ``impurities" of space-time are the D-branes. 
They are 
submanifolds of space-time defined by the condition that open strings can end on them.
They are the stringy resolution of    solutions of GR
that correspond to higher dimensional generalizations of extremal black holes.
These impurities give a rationale for the existence of open string theories. One
can say that while closed strings are the quantum excitations of the smooth part
of space-time, open strings are the quantum excitations of these particular impurities,
the D-branes. 

D-branes are at the core of most of the recent developments in string theory.     
Their most important property is the development of a rank-$N$ 
 nonabelian gauge symmetry
 when $N$ D-branes sit on top of each other.

\subsection{AdS/CFT and non-perturbative strings}

\noindent

The collective dynamics of a single D$p$-brane is a theory of open strings with
endpoints confined to the $(p+1)$-dimensional world-volume. At long
wavelengths this open
 string theory always contains a $U(1)$ gauge multiplet, while an enhanced
$U(N)$ symmetry develops when 
accumulating  $N$
 D-branes at a point in transverse space.  
 At the same time, the gravitational radius of the supergravity solution  
scales as
\begin{equation}
R \sim (g_s N)^{1/4} \;\ell_s\;.
\end{equation}
Hence,
 in the limit $N\gg 1$, $g_s \ll 1$ with $g_s N \gg 1$ the gravitational radius
stays much larger than the string scale and at low energies  GR 
still provides a good description.  
 From the point of view of the $U(N)$ gauge theory, $g_s N =
g^2 N$ is the 't Hooft coupling of the $1/N$ expansion, 
so that the previous limit
corresponds to the large-$N$ expansion of the $SU(N)$ Yang--Mills theory
 with fixed and large 't Hooft coupling \cite{thooftn}.   

The celebrated AdS/CFT conjecture \cite{malda, adscft, magoo}
 states that the large-$N$
dynamics of the gauge theory on the world-volume of the branes is
equivalent to the gravitational description based on the near-horizon
limit of the  supergravity
solution.

 In the cases where the correspondence is well understood, the
gauge theory has an ultraviolet fixed point of the renormalization group,
which defines a conformal field theory (CFT). The corresponding dual geometry
is asymptotic to Anti-de Sitter space (AdS)
 times a compact Einstein manifold: ${\rm AdS} \times K_{\rm E}$.
In the simplest example, we have  a duality between type-IIB strings
on ${\rm AdS}_5 \times {\bf S}^5$ with $N$ units of Ramond--Ramond
flux on the sphere, and the large-$N$ dynamics of ${\cal N}=4$ super
Yang--Mills theory with gauge group $SU(N)$. 

The space-time where the CFT is defined can be characterized as the
conformal boundary of the   AdS gravitational background. For example,
the conformal boundary of ${\rm AdS}_5$ is the conformal class of four-dimensional
Minkowski space.
More explicitly, 
 the correspondence states that the generating functional of CFT
correlation functions equals the quantum partition function of the string
theory with given boundary conditions:
\begin{equation}
\left\langle \exp \left(\int_{\partial X} J\;{\cal O} \right)\right\rangle = 
\exp\left(-I_{\rm eff} \,[\phi \rightarrow \phi_{\partial X} = J]\right)\;.
\end{equation}  
In this expression the gravitational effective action is evaluated as
a function of the boundary values of fields $\phi$ at the boundary
$\partial X$  of the
bulk space-time $X= {\rm AdS} \times K_{\rm E}$. 

Since the CFT
is a standard  quantum field theory without gravity, we may be able
to define it non-perturbatively. In this way, a strong version of the
AdS/CFT correspondence provides a non-perturbative definition of
string theory in certain spaces that are asymptotic to AdS space.

\subsection{Holography and the  entropy test}

\noindent

The AdS/CFT correspondence offers the most explicit realization of the
holographic principle \cite{holo}. According to this principle, 
quantum states associated to a region of
space can be  written in terms of   degrees of freedom on the boundary of this
region. This idea is based on the physics of black holes, 
and in particular the peculiar scaling of
the Bekenstein--Hawking entropy with
the area of the event horizon.
 According
to these ideas, the bulk of space-time is a purely semiclassical concept,
a sort of WKB artefact with a limited range of validity.

The high energy spectrum of finite-energy excitations in a gravity theory is given
by black holes. The 
 largest black holes supported by ${\rm AdS}_{d+1}$ space have an entropy of order  
\begin{equation}
S_{\rm BH} = {A_{\rm Horizon} \over 4\,G_{\rm N}} =
 C\;\left(M\,R \right)^{d-1 \over d}\;, 
\end{equation}
where $M$ is the mass of the black hole and $R$ the radius of curvature of the AdS
space. This is exactly the scaling of the thermal entropy of CFT  
defined on a spatial sphere ${\bf S}^{d-1}$ of radius $R$, provided we identify
$M$ with the CFT energy. Thus, the density of states at
high energy supports the idea of holography: there are enough states in a CFT defined
 on the
boundary of AdS to account for all  finite-energy excitations of gravity. 
 In the cases where the corresponding CFT entropy could be exactly computed, it was found
in complete agreement with the Bekenstein--Hawking formula,
 down to the factor of 1/4, \cite{svafa}. 

This is arguably the most important quantitative test ever made in
string theory, and in some sense it is the first time that the theory meets
an unambiguous numerical check. The importance of this success can hardly be
understated. Its main limitation:  it holds only for certain black holes that
can be regarded as excitations of supersymmetric vacua. Thus, even if the
black holes themselves may or may not be exact BPS states,
 they sit in a Hilbert space that
does admit the action of an exact supersymmetry algebra.
   
\subsection{Fundamental strings as QCD strings}

\noindent

The quantum equivalence between a four-dimensional $SU(N)$ gauge theory and
a string theory with coupling proportional to $1/N$ is an old hypothesis
and one of the conceptual venues towards
 the understanding of quark confinement \cite{thooftn}. What is remarkable in the 
AdS/CFT 
case is the emergence  of 
   a ten-dimensional ``fundamental" string theory,  with gravity and all.
A priori, the QCD string could be anticipated to be some sort of ``effective"
or ``fat" string with a thickness of order $\Lambda_{\rm QCD}^{-1}$ in terms of
the non-perturbative dynamical scale $\Lambda_{\rm QCD}$. Instead, we find a ``thin"
string with gravity in extra dimensions with peculiar negative curvature.
However, the background is such that integrating out the extra dimensions generates
the appropriate degree of non-locality to induce the ``thickness" of the
QCD string in the physical four-dimensional space-time. 

Therefore, the AdS/CFT correspondence offers the first non-trivial example
of large-$N$ string in the sense of \cite{thooftn}. In some cases it was possible
to build models with many properties of QCD, such as confinement, gluon
condensates in the vacuum, etc.  However, all these models are defined
by a soft breaking of supersymmetric and conformal models at some
scale $M_b$. Let $\lambda =g^2 N$
be the value of the 't Hooft coupling at the scale $M_b$. QCD is obtained
by taking $\lambda \ll 1$ so that a large hierarchy of order
\begin{equation}
\log\,\left({M_b \over \Lambda_{\rm QCD}}\right) \sim {1\over \lambda} \gg 1 
\end{equation}
is generated.   
Unfortunately, in all known examples the limit $\lambda \ll 1$ is
technically difficult in terms of the string theory. Standard approximate
methods, based on supergravity, 
 only apply to $\lambda \gg 1$ and in this strong coupling regime
one has
\begin{equation}
\log\,\left({\Lambda_{\rm QCD} \over M_b}\right) \sim \lambda \gg 1\;, 
\end{equation}
violating the  scaling of an asymptotically free theory.  
Thus, a constructive procedure exists to approach QCD on the string
side, starting with some well defined models;  unfortunately
the result stays out of calculational reach. Extrapolations from
$\lambda\gg 1$ have mostly heuristic value, since large-$N$ phase
transitions in $\lambda$ can be expected on general grounds.  

\subsection{Phenomenology}

\noindent

String  model building has long known good  
approximations  of the (supersymmetric)  Standard Model,
 including features such as the gauge group, generations
and chiral representations.  The pool of available options was 
considerably enlarged by  the  consideration of 
geometrical models of the form ${\bf R}^4 \times K^*$, where the asterisk
stands for the possibility of
decorating the compact manifold with
 various ``impurities" in the form of branes and fluxes. 

This enhanced diversity comes  at the price of removing many of  
the ``model-independent" predictions of old models based on the weakly
coupled heterotic
string on Calabi--Yau manifolds.
One of the most quantitative such predictions  was
the relation between the Grand Unification  scale $M_{\rm GUT}$ and Newton's
constant: 
\begin{equation}
G_{\rm N} > {\alpha_{\rm GUT}^{4/3} \over M_{\rm GUT}^2}\;,
\end{equation}
where $\alpha_{\rm GUT} \sim 1/25$ is the value of the gauge couplings at
the unification scale. The bound on $G_{\rm N}$ is a consequence of requiring
the string theory to be weakly coupled, $g_s <1$, and
 comes out too large by
a factor of about $400$, which must be blamed on  threshold corrections.
In the last few years it was recognized that localizing the SM gauge
interactions on singularities of $K^*$, notably branes of various kinds,
one could remove this constraint 
\cite{horavaw}. The
general tree-level formula for the string mass scale
\begin{equation}
M_s \sim \left({g_s^2 \over G_{\rm N} \;{\rm Vol}\,(K^*)}\right)^{1/8}
\end{equation}
allows us to lower $M_s$ down to a few TeV, provided we increase 
the size of the compact manifold. These large extra dimensions are
transverse to the branes that confine the SM fields, so that they are
largely invisible to SM processes. This is the much studied scenario of
{\it large extra dimensions} 
 \cite{ahdd}. One can in principle build models with a string scale
anywhere in the range 
\begin{equation}
{\rm few} \;\;\;{\rm TeV} \;< M_s < 10^{18} \;\;{\rm GeV}
\;, 
\end{equation}
with better and better qualitative  matching of the SM at low energies (see for example
\cite{lust} for a recent summary). 
 In fact, the relevant geometrical
parameter of $K^*$ is not necessarily its volume, it can also be the
radius of (negative) curvature, leading to the {\it warped} scenario of 
\cite{rs, giddings}. 

The emphasis on chirality
 reduces the choices to models with ${\cal N}=1$ supersymmetry.
In perturbation theory in a given string theory, such models have exactly massless
moduli and exact supersymmetry. Thus, the perturbative approximation yields a
moduli space of vacua with
 ${\cal N}=1$ supersymmetry in four dimensions, 
 in spite of the expectation that {\it exact} vacua should be
isolated.  

The moduli fields appear in four dimensions as gravitationally coupled scalars,
and they 
 are very constrained experimentally. In fact, they are the major
embarrassment for string models that approximate qualitative features of the
Standard Model.
To date,
no realistic vacua without moduli could be constructed explicitly.
Most of the work has proceeded by trying to lift
the moduli and break supersymmetry
at the same time, all in the perturbative shores of the moduli space where we
can justify the calculations. Thus, the so-called ``moduli problem" has been tied
to the problem of supersymmetry breaking, in part  for technical reasons.
The lack of a satisfactory solution of these problems stands as the main obstacle
in rendering string phenomenology predictive.

\section{A selection of recent developments}

\noindent

With no claim to completeness, we will mention some of the most
significant trends of theoretical research in string theory. Leaving aside
important results in  mathematical physics (for example \cite{dv}) we will
focus here on the more physically-motivated questions.  
 
On the one hand, there is important activity  
in  problems posed by the AdS/CFT correspondence,
both in its application to  quantum gravity and to the problem
of the QCD string. On the other hand, we have witnessed a revival of
the study of time-dependent backgrounds with applications to
cosmology as well as important progress in the classic problem of
moduli stabilization.

\subsection{Towards the QCD string}

\noindent

A very active area of research is the ongoing effort to bring the
AdS/CFT models closer to real QCD by gradually lifting the constraints
of conformal symmetry and supersymmetry.

As explained above,   AdS/CFT models with soft breaking at scale
$M_b$ can be studied in the supergravity approximation in an expasion
in powers of $M_b /\Lambda_{\rm QCD}$. In order to invert this expansion parameter
and approach the physical regime of pure non-supersymmetric Yang--Mills
theory, one must solve the string theory exactly in the $N\rightarrow \infty$ limit.

Although this feat remains beyond our present capabilities, interesting
progress has been achieved recently in certain kinematical limits.

 One
possibility is to study the original AdS/CFT model in the limit of
large R-charge \cite{bmn}. Out of the global $SO(6)$ R-symmetry of
 the ${\cal N}=4$ theory we may select a $U(1)$ subgroup
and consider the limit of large charge. In the gravitational description,
this introduces an infinite boost of the ${\rm AdS}_5 \times {\bf S}^5$
geometry along an equator of ${\bf S}^5$. In this limit the background  
simplifies and we can solve exactly the tree-level string
theory in the light-cone gauge. Thus we can
extend the holographic correspondence beyond the BPS limit, provided
we  zoom into this sector of the total Hilbert space. The result is a
rich generalization of AdS/CFT with interesting questions
about the rules of holography and the role of string field theory.

Another interesting kinematical limit is that of large spin in the
physical space. One considers the gauge theory on a spatial 3-sphere
and takes the limit of large angular momentum $J$ on an equator of ${\bf S}^3$.
 On the AdS side, it is then possible to
 identify special solitonic string states dual to
operators satisfying \cite{gkp}
\begin{equation}
\Delta - J \sim \log\;J\;,
\end{equation}
where $\Delta$
stands for the anomalous dimension of the operator. The left-hand side
of this equation is nothing but  the {\it twist} of the operator,
in the language of deep inelastic
scattering. In fact, the logarithmic behaviour is a famous consequence
of asymptotic freedom in perturbation theory \cite{gw}.

 It is rather
intriguing to see this logarithm arising here from a purely geometrical
calculation. It shows that focusing on special operators of large quantum
numbers one can hope to 
bridge the gap between the weak and strong 't Hooft coupling.  

The world-sheet description of these
  kinematical limits is  related to certain two-dimensional integrable systems, 
a source of much recent interest (see \cite{inte} for a  summary of the growing literature
on the subject), in striking analogy with
known results in high-energy QCD \cite{lipatov}.   A different connection between
perturbative gauge theory and string theory was recently uncovered in \cite{twistors}.

\subsection{The question of background independence}

\noindent

One  crucial lesson of current non-perturbative definitions of string
theory is their dependence on asymptotic boundary conditions. For
asymptotically AdS spaces we can define appropriate boundary
conditions that specify a Hamiltonian. For asymptotically flat
spaces we are just able to define an S-matrix that might be calculable
in principle through a limit of AdS/CFT or perhaps the matrix theory of
\cite{bfss}.
 What could be the analogous structures relevant to space-times
with closed spatial sections and/or cosmological singularities
 remains a  mystery. For many years it was assumed
that string field theory 
would hold the answer by providing a non-perturbative,
 background-independent formulation of string theory. However, the
developments centred about the realizations of holography (matrix
theory and the AdS/CFT correspondence) severely question this hope. In fact,
most evidence   based on existing models tends to discourage the idea
of background independence.

The simplest  example of a background not falling in the understood
categories is  de Sitter space, the maximally symmetric space of constant
positive curvature.   It breaks supersymmetry and cannot be recovered as
a smooth compactification of higher-dimensional supergravity \cite{nogo}.  
Allowing ``impurities" in the compact manifold, such as D-branes and fluxes,
it seems possible to construct {\it metastable} vacua with positive
cosmological constant, i.e. metastable de Sitter bubbles \cite{kklt}.

Thus, one possibility is that de Sitter space can only be defined as
a metastable resonance in the S-matrix of an asymptotically flat, supersymmetric,
vacuum \cite{reson}, but there are at least two  other, more radical proposals.

 One is the  so-called
dS/CFT correspondence of \cite{dscft}, a sort of analytic continuation
of AdS/CFT with a different physical interpretation in which  cosmological
time is identified with a renormalization group flow between conformal
fixed points.  Yet another one uses a  radical interpretation of holography
to claim that quantum de Sitter space has a finite-dimensional Hilbert
space with most states localized at the observer's event
 horizon \cite{banks}. In
this proposal the cosmological constant is an input related to the dimension
of the Hilbert space rather than a calculable parameter.    

Such a diversity of proposals that are well motivated, and yet so different
at the conceptual level, show how fascinating this problem is,
but they also reveal the primitive stage of our understanding. 

\subsection{Time-dependent backgrounds and cosmological singularities}

\noindent 

Although string-inspired ideas soon found their way into cosmology \cite{cbs}, 
time-dependent backgrounds in string theory 
have been comparatively less studied 
beyond the supergravity approximation. 
 In perturbation theory, the corresponding world-sheet
conformal field theories are difficult to analyse.
 Despite these problems, there exist interesting
cosmological backgrounds, which are based on non-compact coset models such as  
the classic model of \cite{nappiw}. Many physical aspects of these space-times  
have been studied recently \cite{rab}, although the computation of
the S-matrix  beyond the one-particle scattering is still a notorious challenge.  

Time-dependent orbifolds introduced in \cite{horos}
 are more amenable to analytic treatment
and were extensively studied as toy models of pulsating universes
 (see for example the recent review
 \cite{torb}). 
 The main result of these studies is negative, in
the sense that back-reaction gets out of control of perturbation theory
near the singularity. Very general arguments support the idea that
these cosmological singularities are ultimately as hard as generic
black hole singularities \cite{hpol}.  

At a non-perturbative level, all the dilemmas afflicting the quantum
mechanics of de Sitter space come back, in an even more agressive
incarnation, since the singularities may deprive us from smooth asymptotic
regions, where the specification of the Hilbert space could be easier.  

In general, the resolution of space-like singularities in string theory
is still uncharted territory. The great progress in the resolution of
static (i.e. timelike) singularities is largely a consequence of
supersymmetry and duality, while the big bang of a FRW model or
the singularity of a black hole feature a maximal violation of supersymmetry. 

Nevertheless, important lessons for cosmological singularities
lie hidden in the AdS/CFT correspondence. 
Since large AdS  black holes can be realized  as thermal states of the CFT,
it should be possible to extract information about the internal singularity
from the thermal correlation functions of appropriate operators. The difficulty
in doing so is our poor understanding of the holographic map beyond very
symmetric or generic states.  
This is a fascinating (albeit difficult) set of problems
whose exploration is only beginning
 \cite{maldab}. 

Currently, a large effort is being devoted to the understanding of the simpler
problem of time-dependent open-string backgrounds. These can be interpreted
as dynamical processes involving unstable branes (D-brane decay) or
systems of branes (D-brane anti-D-brane annihilation). These systems  have
even been proposed as the basis of some exotic cosmological models in the context
of the large extra dimensions scenario  \cite{ekp}. See  \cite{maldak} for
a recent summary of applications to inflationary models.    

Perturbatively in $g_s$, these processes are determined by boundary
perturbations of the world-sheet conformal field theory \cite{sen}. Conversely,
on the world-volume of the branes we have the dynamics of a tachyonic 
mode rolling down a potential. This is a characteristic problem of
open string field theory and with this motivation it has been much studied. 
Recently, 
  the crucial issue of back-reaction was tackled
in the context of two-dimensional toy models \cite{cuno}.

\subsection{The Landscape} 

\noindent 

As pointed out above, the existence of massless moduli fields coupled
gravitationally stands out as an unphysical feature of supersymmetric
string models of low-energy
 phenomenology. For semi-realistic ${\cal N}=1$ models,
such a defect is presumably an artefact of perturbation theory. Yet,
the problem of {\it moduli stabilization} stands as a classic difficulty
in rendering any model quantitative.  

Upon supersymmetry-breaking, the problem becomes more complicated, and  
 ties up  with  the thorny issue of the cosmological constant. Typical effective
potentials for moduli, based  for example on scenarios of gaugino condensation,
show runaway behaviour  unless one fine-tunes the dynamics to
achieve a stable vacuum at weak coupling. In this case, one  finds
the moduli masses  and the cosmological constant controlled by the supersymmetry
breaking scale $M_{b} \sim 1$ TeV. Even postponing the problem of the
vacuum energy, moduli masses around the TeV scale cause notorious
problems to the standard theory of nucleosynthesis \cite{modulip}.  

Recently, significant progress was achieved in the purely technical problem
of stabilizing moduli. In fashionable models of the form ${\bf R}^4 \times
K^*$, where the compact space $K$ is decorated with branes, orbifold
singularities and trapped magnetic fluxes, there are a huge number of
discrete choices for $K^*$, and it was found that the effective potential
on ${\bf R}^4$ depends on these quantum numbers in an intrincate way (see
the contribution of T. Taylor to this conference for more details).

For example, consider
 $N$ units of magnetic flux
\begin{equation}
\oint_\Sigma F = N
\,,\end{equation}
 trapped on a submanifold $\Sigma$
inside $K^*$. In normal models there are dozens of independent fluxes of
this type. The contribution to the effective potential scales like
\begin{equation}
V_{\rm flux} \sim \int_{K^*} |F|^2 \sim {{\rm Vol}\,(K) \;N^2 \over {\rm Vol}
\,(\Sigma)^2}\;.
\end{equation}
On the other hand, $N'$ wrapped branes on a cycle $\Sigma'$ contribute
\begin{equation}
V_{\rm brane} \sim N'\;T_{\rm brane}  \;\cdot\, {\rm Vol}\,(\Sigma')\;.
\end{equation}
Again, a typical scenario may have hundreds of such independent wrapping
modes. Combining many fluxes and branes and including gravitational corrections, 
one can derive an effective potential that fixes most moduli for
each choice of the set of discrete quantum numbers $N_i$.  

In this method,  
 one literally stabilizes the internal manifold $K^*$
in a ``mechanical" fashion, equilibrating tension force from the wrapping
with magnetic repulsion from the trapped fluxes. To be precise, the
modulus corresponding to the overall size, ${\rm Vol}\,(K^*)$, remains
unfixed in these models, and one must invoke other mechanisms,  
such as the old gaugino condensation, to complete the job \cite{kklt}.

One interesting aspect of these methods is their versatility, potentially
applying to many model-building scenarios. For example, one can think
of fixing the moduli with masses $m_\phi \gg {\rm TeV}$, thus alleviating
the cosmological moduli problems. 

However, perhaps the most striking aspect of this scenario is its new angle
on the cosmological constant problem. Simplifying things a bit, the
 contribution of fluxes to
the vacuum energy depends on the discrete numbers $N_i$ as
\begin{equation}
V_{\rm min} = \Lambda_b + \sum_{i=1}^{n_f} C_i \,N_i^2\;,
\end{equation}
where $\Lambda_b$ stands for the contribution from other sources and
$n_f$ is the number of relevant independent fluxes (easily of $O(100)$). 
Assuming
that $\Lambda_b$  is negative, the  vacua with cosmological 
constant in the physical range,  $\Lambda_{\rm ph} \pm \delta \Lambda$,  
are the solutions of the discrete equation
\begin{equation}
|\Lambda_b | + \Lambda_{\rm ph}  - \delta \Lambda <
 \sum_{i=1}^{n_f} C_i \;N_i^2 < |\Lambda_b | + \Lambda_{\rm ph}  + \delta \Lambda   
\;.
\end{equation}
The number of vacua in the  band of width $2\delta \Lambda$  
grows exponentially with $n_f$; we have a  quasicontinuous spectrum of
vacua that is known as the {\it discretuum} \cite{boussopol}. 

In this scheme, we are virtually garanteed of finding an astronomical number of vacua
with cosmological constant within acceptable limits. Recent estimates 
yield exponentially large numbers in the range of $10^{100}$ \cite{doug}.
 In principle, a small fraction
of these will have other desirable features, such as large mass hierarchies and
correct particle content, and  one can hope that the SM will appear ``in the
list". However, with such large numbers of vacua involved, one must wonder
whether the scheme is at all testable, even in principle.   

It should be mentioned that these considerations are based on the somewhat ill-defined
concept of effective potential over the perturbative moduli space (the string {\it landscape}
of \cite{reson}). This approach has potential caveats \cite{banksc, dine} and it remains 
unclear what will be
 the precise mathematical status of the ``discretuum" of vacua, beyond
the supersymmetric subset. 

In general, this type of  ``landscape phenomenology" represents a radical departure from
traditional thinking about naturalness problems. When embedded into a scenario of
eternal inflation \cite{reson}, the landscape can  address fine-tuning problems
by a contingent choice of vacuum out of a huge discrete set of possibilities. In such a
context, it is no longer clear whether a particular  small parameter
 has a purely  environmental value (such as the cosmological constant),
 or whether it could be explained by a concrete  
mechanism (such as the proton mass). A more detailed
 discussion of
the landscape phenomenology appears in  \cite{dine, dined}

\section{Concluding remarks}

\noindent

Our survey shows that enormous progress was achieved in elucidating the
conceptual status of string theory as a model of quantum gravity. There 
is a global picture of models with extended super-Poincar\'e symmetry and
a fairly explicit   non-perturbative formulation of the theory
on asymptotically AdS spaces. This formulation conforms to the general
ideas of holography and the successful calculation of the Bekenstein--Hawking entropy
for certain black holes stands as the main quantitative test of these  results.   

The current frontier of development lies in the extension of these ideas to
 non-supersymmetric
space-times, notably backgrounds with cosmological interpretation, a notoriously
hard challenge.   
 This is arguably the area of string theory in most urgent need for improvement,
 because even the simplest of examples, de Sitter space, poses a formidable
theoretical challenge.

 Of course, de Sitter space is also quickly becoming  
a phenomenological urgency, given the apparent measurement of a strictly 
 positive 
cosmological constant \cite{lambda} and the mounting evidence in favour of an early
inflationary era in our Universe \cite{wmap}. On the positive side, this means that 
string theory and quantum gravity could be closer than expected to
experimental tests.  

The AdS/CFT correspondence also provides the first examples of large-$N$
gauge strings in four dimensions, with non-trivial dynamical properties such
as confinement. The successful lifting of the constraints of supersymmetry and conformal
symmetry remains the main obstacle in the approach to real QCD, a difficult but
extremely important  problem.  

The reformulation of  the unification paradigm in terms of strings 
provides a global
framework for virtually all    past scenarios of physics beyond the SM.  
In recent years, thanks to the versatility of D-branes and the understanding of
duality symmetries,   the number of
quasi-realistic models has increased considerably, at the price of losing some
old ``model-independent" predictions.  It is now possible to entertain
many model-building possibilities, some 
 with fundamental scale as low as a few TeV, changing the traditional
 perspective on  
the mass hierarchy problems and opening new exciting experimental prospects.
 The  stabilization of moduli in a physically
acceptable way remains as a major  problem in which we are seeing considerable
progress. The picture of a {\it discretuum} of vacua gradually emerges, to the
discomfort of many, who would like a more predictive scenario.
Generally speaking, the rigorous  existence and properties of a landscape of vacua
becomes the main question of principle to be addressed in this context.  

One physical property pervades the whole theoretical building of string theory
as we know it: supersymmetry. A radical but well motivated view would hold that
supersymmetry is not just an offspin of string theory, but rather lies at its very
foundation.  Although supersymmetry at the TeV scale is not a solid prediction,
finding experimental evidence in its favour would be of the utmost importance for
string theory. 

\subsection*{Acknowledgement} 

\noindent

I would like to thank the organizers of EPS2003, and especially
 Prof. Ch. Berger, for giving me
 the oportunity to enjoy such an interesting conference. 
 
%

%
%
%
%

\end{document}